\documentclass[twocolumn,preprintnumbers,amsmath,amssymb,longbibliography]{revtex4-2}

\usepackage{graphicx}
\usepackage{dcolumn}
\usepackage{float} 
\usepackage{bbm}

\usepackage{setspace}
\begin{document}

\title{Dynamical Symmetries and Symmetry-Protected Selection Rules  in Periodically Driven Quantum Systems}

\author{ Georg Engelhardt$^1$    }

\author{ Jianshu Cao$^{2} $}
\email{jianshu@mit.edu}

\affiliation{%
$^1$Beijing Computational Science Research Center, Beijing 100193, People's Republic of China\\
$^2$Department of Chemistry, Massachusetts Institute of Technology, 77 Massachusetts Avenue,
Cambridge, Massachusetts 02139, USA
}

\date{\today}

\pacs{%42.50.Dv,	%Quantum state engineering and measurements
      %74.50.+r,	%Tunneling phenomena; point contacts, weak links, Josephson effects
      %05.60.Gg,		%Quantum transport
      %03.65.Vf,		%Topological phases (quantum mechanics)
      %73.63.Kv,		%Quantum dots
      %73.23.Hk		%Coulomb blockade; single-electron tunneling 
      %73.23.-b		%Electronic transport in mesoscopic systems
      %03.67.Lx,	%Quantum computation
      %05.60.-k		%Transport processes
      %05.45.Jn      %High-dimensional chaos
}

\begin{abstract}
	In recent experiments,  the  light-matter interaction has reached the ultrastrong coupling limit, which can give rise to dynamical generalizations of spatial symmetries in periodically driven systems. Here, we present a unified framework of dynamical-symmetry-protected selection rules  based on Floquet response theory.
	Within this framework, we study rotational, parity, particle-hole, chiral, and time-reversal symmetries and the resulting selection rules in spectroscopy, including symmetry-protected dark states (spDSs), symmetry-protected dark bands (spDBs), and symmetry-induced transparency (siT). Specifically,  dynamical  rotational and parity symmetries establish spDS  and  spDB conditions. Particle-hole symmetry introduces spDSs for symmetry-related Floquet states and also   a siT at quasienergy crossings. Chiral symmetry and time-reversal symmetry alone do not imply  spDS conditions but can be combined to define a particle-hole symmetry. These symmetry conditions arise from  destructive interference due to the synchronization of  symmetric quantum systems with the periodic driving. Our predictions reveal new physical phenomena when a quantum system reaches the strong light-matter coupling regime, which is important for  superconducting qubits, atoms  and molecules in optical fields or  plasmonic  cavities,  and optomechanical systems.
\end{abstract}

\maketitle

\allowdisplaybreaks

%-----------------------------------------------------------------------------
\textit{Introduction.}
 Over the last few decades, the light-matter interaction strength has been pushed to the ultrastrong coupling regime in optomechanical systems~\cite{Fang2017}, quantum dots,  atoms and molecules in optical or plasmonic cavities~\cite{Yin2020,Li2020,Pino2015,Shalabney2015, Sukharev2017}, and superconducting quantum circuits~\cite{Pietikaeinen2017,Forn-Diaz2019,Kockum2019}. 
As standard nonlinear perturbation theory~\cite{Mukamel1995} becomes  unfeasible under these conditions,  Floquet response theory has been developed recently, describing  systems that are subject to a  strong but time-periodic driving field (of frequency $\Omega$), and a weak but arbitrary probe field~\cite{Kohler2018,Gu2018,Cabra2020,Engelhardta}. For a monochromatic probe of frequency $\omega_{ p}$, system observables generate response frequencies $\omega_{ p}+ n\Omega$ termed \textit{Floquet bands}~\cite{Kumar2020}. 

Spatial symmetries give rise to appealing physical properties. Inversion symmetry results in selection rules for dipole transitions;  particle-hole, chiral,  and time-reversal symmetries establish the so-called periodic table, a classification scheme for topological insulators~\cite{Altland1997,Langbehn2017,Kawabata2019},   and symmetries have an essential impact on transport properties~\cite{Duan2020,Thingna2020,Wu2013a,Chuang2016,Thingna2016,Engelhardt2019a,Fleming2009}.
For periodically driven systems,   these spatial symmetries can be generalized to dynamical symmetries  that can give rise to a generalized periodic table for topological insulators~\cite{Roy2017,Peng2019} and new control mechanisms ~\cite{Engelhardt2016,Engelhardt2017a,Gomez-Leon2012,Yuen-Zhou2015,Yan2019,Yan2018,Chinzei2020,Buca2019}.  Dynamical symmetries have been used to control the coherent destruction of tunneling effect~\cite{Grossmann1991} and induce selection rules for
high harmonic generation~\cite{Bavli1993,Alon1998,Alon2000,Alon2002}.

\begin{table}[t]
	\caption{Overview of the spectroscopic signatures of dynamical rotational symmetry (RS), particle-hole symmetry (PHS), parity symmetry (PS), chiral symmetry (CS), and time-reversal symmetry (TRS). The signatures include symmetry-protected dark states (spDSs), symmetry-protected dark bands (spDBs), symmetry-induced transparency (siT), and accidental dark states (aDSs). The rightmost column lists example models.}\label{tab:overview}
	\begin{tabular}{c|c|c}
		Symmetry &   Effect   & Example \\
		\hline
		\hline
		RS   &    spDS    &   benzene ring  (Fig.~\ref{figSpectraBenzene}) \\
		    &   spDB      &  \\
		\hline
		PS   &    spDS    &   two-level system (Fig.~\ref{figSpectraSpinBoson}) \\
		&   spDB      &  \\
		\hline
		PHS   &   spDS  &  dimer (Fig.~\ref{figSpectraDimerModel})  \\
		\hline
		2 $\times$ PHS   &   siT     & two-level system (Fig.~\ref{figSpectraSpinBoson})\\
		\hline
		TRS &   none     &         \\
		\hline
		CS &   none        &    \\
		\hline
		 none  &  aDS   &  all  (Figs.~\ref{figSpectraBenzene},\ref{figSpectraDimerModel}  ,\ref{figSpectraSpinBoson})
	\end{tabular}
\end{table}

In this Letter, we introduce a unified conceptual framework of selection rules based on general dynamical symmetries of periodically driven quantum systems as described by Floquet
response theory~\cite{Engelhardt2020}.  Physically, the synchronization of symmetric quantum systems with the periodic driving gives rise to  destructive interference  effects in Floquet space and thus to forbidden transitions between Floquet states. This set of forbidden transitions defines the symmetry-protected selection rules that are robust against symmetry-preserving parameter variations.  Specifically, there are four types of forbidden transitions ordered in  increasing degree of complexity: (i) accidental dark states (aDSs), appearing for a 
specific combination of system parameters; (ii) symmetry-protected dark states (spDSs),
which refer to the symmetry-protected absence of a complete transition line similar to symmetry-protected excitations of topological band structures; (iii) symmetry-protected dark bands (spDBs), which refer to the absence of a complete Floquet band due to a combination of spDSs;  (iv) symmetry-induced transparency (siT), which refers to the vanishing transition intensity at the degeneracy of quasienergies.  Except for the aDS, which is
not symmetry related,  we establish symmetry-protected selection rules for important dynamical symmetries, which are classified in   Table~\ref{tab:overview}.

\textit{Floquet response theory.}
We apply a semiclassical approach based on the general Hamiltonian
\begin{equation}
\hat H(t) =  \hat H_0(t) +  \int_{0}^{\infty}  d\omega  \left[ \lambda  \hat  V \left( \hat a_{\omega}   + \hat a_{\omega}^{\dagger}  \right)  + \omega  \hat a_{\omega}^{\dagger}   \hat a_{\omega}\right]   ,
\label{eq:fullHamiltonian}
\end{equation}
where $\hat H_0(t) = \hat H_0(t + \tau)$ describes a system driven by a periodic classical electromagnetic field of frequency $\Omega= 2\pi/\tau$. The probe field  is given by a continuum of  photonic operators $a_\omega^{\dagger}$ with frequencies $\omega$, which are coupled via the dipole transition operator $\hat V$ with strength $\lambda$ to the driven system. The physical properties of $\hat H_0(t)$ are determined by the Floquet  equation
\begin{equation}
\left[ \hat H_0(t)-i\frac{d}{dt}   \right] \left| u_\mu (t)\right>   =   \epsilon_\mu \left| u_\mu (t)\right> ,
\label{eq:FloquetEquation}
\end{equation}
where $\left| u_\mu (t)\right> = \left| u_\mu (t+\tau)\right>  $ and  $ \epsilon_\mu$  are  the corresponding Floquet states  and  quasienergies that generalize the concept of eigenstates and eigenenergies of   time-independent systems.
It is implicitly assumed that the driven system is weakly dissipative such that,  for long times, it approaches the stationary state
\begin{equation}
\rho(t) = \sum_{\mu}  p_\mu \left| u_\mu (t)\right> \left< u_\mu(t) \right|  ,
\label{eq:stationaryState}
\end{equation}
which is diagonal in the Floquet basis and thus synchronizes with the driving  $\rho(t) =\rho(t+\tau) $. Equation~\eqref{eq:stationaryState} is consistent with the Floquet-Redfield equation~\cite{Kohler1998,Blattmann2015,Shirai2015,Engelhardt2019} describing periodically driven open quantum systems. In the model calculations, we assume the special distribution  $p_\mu \propto  e^{-\beta \epsilon_\mu}$, i.e., a Floquet-Gibbs distribution, but all our predictions  hold even if the Floquet-Gibbs distribution breaks down~\cite{Engelhardt2019}. Strongly dissipative systems could be addressed by generalizing our approach to non-Hermitian Hamiltonians~\cite{Kawabata2019} or via the polaron transformation~\cite{Xu2016}.

The interaction of $\hat H_0(t)$ with the probe field is treated using the input-output formalism and a perturbation expansion for small $\lambda$. The input field consists of a bichromatic probe field (of frequencies $\omega_{ p,1} $ and $\omega_{ p,2 } = \omega_{ p,1} + n \Omega $, integer $n$).  As  shown  separately~\cite{Engelhardta}, the intensity change of the output field at frequency $\omega_{ p,2 }$   proportional to the  coherence $ \left<\hat a_{\omega_{ p,2}}^\dagger  \hat a_{\omega_{ p,1}}\right> $  is given by
$
\Delta I_{\rm coh}(\omega_{ p,2}  )=     -i\tilde \chi_n (\omega_{ p,1})  \left<\hat a_{\omega_{ p,2}}^\dagger  \hat a_{\omega_{ p,1}}\right> + \text{c.c.}
$, where the susceptibility $\tilde \chi_{n}(\omega_{ p,1})$ can be evaluated using Floquet response theory
and reads
\begin{eqnarray}
\tilde \chi_{n}(\omega_{ p,1}) &=& i\lambda^2 \sum_{\nu,\mu,m} \frac{ V_{\nu,\mu }^{(-n-m)}  V_{\mu ,\nu}^{(m)}\left(p_\nu -p_\mu \right) }{ \epsilon_\mu -\epsilon_\nu +m\Omega  - \omega_{ p,1} - i \gamma_{\nu,\mu }^{(m)}    } 
\label{eq:dynamicSusceptibiltyFrequency}.
\end{eqnarray}
The index $n$ denotes  the Floquet band, which describes nonelastic scattering of the probe field, and 
the dynamical dipole matrix elements read
\begin{equation}
V_{\lambda,\mu }^{(n)} = \frac{1}{\tau} \int_{0}^{\tau}  \left< u_\lambda (t)  \right| \hat V    \left|u_\mu  (t)\right> e^{ - i n \Omega t}dt .
\label{eq:dynamicalDipolElements}
\end{equation}
The parameters $\gamma_{\nu,\mu }^{(m)}$ have been added phenomenologically and denote dephasing rates.

 \textit{Unified conceptual framework of dynamical-symmetry-protected selection rules}.
We consider  the following class of symmetry operations~\cite{Roy2017}:
\begin{equation}
\hat \Sigma \left[ \hat H_0(t_S+ \beta_S t)-i\frac{d}{dt} \right] \hat \Sigma^{-1}  =   \alpha_S \left[ \hat H_0(t)-i\frac{d}{dt} \right],
\label{eq:symmetryRelationFloquetSpace}
\end{equation}
where
 $\hat \Sigma$ is  a time-independent spatial operator.
By specifying $\hat \Sigma $, $t_{ S}$, and $(\alpha_S,\beta_s = \pm1)$,  one can define a set of dynamical symmetries. Applying  Eq.~\eqref{eq:symmetryRelationFloquetSpace} to the Floquet equation Eq.~\eqref{eq:FloquetEquation}, one can identify relations between  Floquet states $\mu$ and $\mu'$: 
\begin{equation}
 \left|u_{\mu'} ( t) \right> =\pi^{(S)}_{\mu} \hat \Sigma \left|u_{\mu} (t_S+ \beta_S t) \right>     .
\label{eq:symmetryRelationFloquetStates}
\end{equation}
These relations can be used to evaluate the dynamical dipole elements in Eq.~\eqref{eq:dynamicalDipolElements}. Imposing an invariance condition for the transition dipole operator $\hat \Sigma^{\dagger} \hat V  \hat \Sigma = \alpha_V^{(S)} \hat V $ and  using Eq.~\eqref{eq:symmetryRelationFloquetStates}, we investigate symmetry-protected selection rules for rotational, parity, particle-hole, chiral and time-reversal symmetries.  

Within the unified framework, we can establish symmetry-protected selection rules that are robust against symmetry-conserved variations and unique for  strong light-matter interactions. 
 Among others, we investigate dark states, which are defined by the condition $V_{\nu ,\mu} ^{(m)}  = 0$, such that the corresponding resonances in
Eq.~\eqref{eq:dynamicSusceptibiltyFrequency} vanish.   This condition not only generalizes the dark state condition in
the standard response theory to the strong-coupling regime for $n =  0$ but also introduces
distinct dark states effects for $n\neq 0$. All selection rules are a consequence of destructive interference due to the synchronization of the system state with the periodic driving: (i) The dark state condition can be fulfilled by special combinations of parameters, which we denote as an aDS, or (ii) as a consequence of a symmetry, which we denote as a spDS.  (iii)  An entire Floquet band can vanish because $\tilde \chi_{n}(\omega_{p})=0$ for specific $n$, which we denote as a spDB.  (iv) By analyzing the susceptibility in terms of Eq. \eqref{eq:symmetryRelationFloquetStates}, we  establish the condition for the siT, which is due to a destructive interference of two transitions with $V_{\nu ,\mu} ^{(m)}  \neq 0$.

\begin{figure}[t]
	\includegraphics[width=0.9\linewidth]{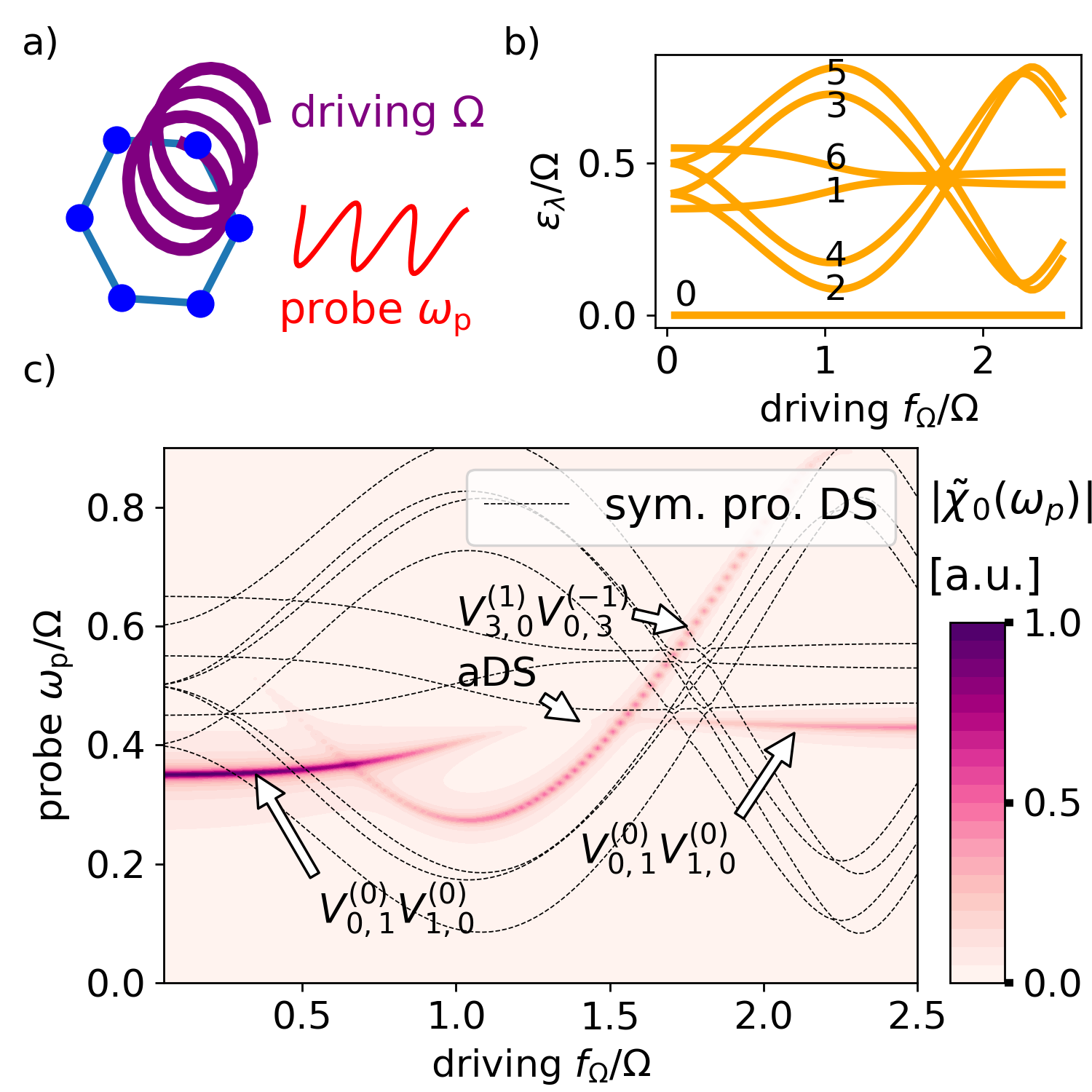}
	\caption{(a) Benzene  driven by circularly polarized light propagating perpendicular to the ring plane. The probe field is polarized perpendicular to the plane so that it does not destroy the sixfold dynamical rotational symmetry. (b) Quasienergies of  benzene    for the tunneling constant $J_0 =0.05\Omega$ and on-site energy $E_0=0.45 \Omega$. (c) Susceptibility   $\tilde \chi_0(\omega_{ p})$ (color gradient). Rotational spDSs are marked  by dashed lines.  One  transition vanishes at the location of the aDS. The dephasing rates in all figures are $\gamma_{\nu,\mu }^{(m)} = 0.001 \Omega$. }
	\label{figSpectraBenzene}
\end{figure}%

\textit{Rotational symmetry.}
With $\alpha_S= \beta_S =1$,  a unitary $\hat \Sigma = \hat R$, and $t_S= t_R =\frac \tau N $ with a positive integer $N$, Eq.~\eqref{eq:symmetryRelationFloquetSpace} defines a dynamical rotational symmetry~\cite{noteRotationalSymmetry} that gives rise to the eigenvalue equation
$
\left| u_\mu (t) \right> = \pi_\mu^{(R)}  \hat R \;  \left| u_\mu (t+t_R) \right> 
\label{eq:parityEigenvalueEquation}
$
with eigenvalues $\pi_\mu^{(R)} = e^{i 2\pi m_\mu /N  }$ and integer $m_\mu =\left\lbrace 0,N-1 \right\rbrace  $. 
 As shown in detail in the Supplemental Material~\cite{supplementals}, for a dipole transition operator with $ \hat R^\dagger \hat V \hat R= \alpha_V^{(R)} \hat V$ and $\alpha_V^{(R)}=\pm 1$,  the dynamical rotational symmetry establishes  a sufficient condition for spDSs:
\begin{equation}
\hat	V_{\nu ,\mu} ^{(m)} \propto 
\begin{cases}
1   &\text{if}\; e^{i \frac{2\pi}{N}\left( m_\mu-m_\nu+ m \right) }\alpha_V^{(R)}  =1, \\ 
0  & \text{else} .
\label{eq:symmetryProtectedDarkStates}
\end{cases}
\end{equation}
Applying Eq.~\eqref{eq:symmetryProtectedDarkStates} to evaluate the susceptibility in Eq.~\eqref{eq:dynamicSusceptibiltyFrequency}, we find
\begin{eqnarray}
\tilde \chi_{n}(\omega_{p}) =
 \begin{cases}
1   &\text{if}\; e^{i \frac{2\pi}{N}n } =1, \\ 
0  & \text{else} ,
\end{cases} \label{eq:floquetBandSelectionRule}
\end{eqnarray}
which is the condition for the complete disappearance of Floquet band $n$, i.e., a spDB.    Physically, this effect appears as  the stationary state Eq.~\eqref{eq:stationaryState} synchronizes with the driving field such that the density matrix adopts the dynamical rotational symmetry, i.e., $\rho(t+n/N\tau)  = \hat R^{n} \rho(t) \hat R^{\dagger n} $.

As an example, we consider a benzene ring  driven by circularly polarized light sketched in Fig.~\ref{figSpectraBenzene}(a), which is described by a tight-binding Hamiltonian:
\begin{eqnarray}
\hat H_0(t) &=& \sum_{j,j' =1}^{6}  J_{j,j'} \left|e_j \right> \left<e_{j'} \right|   
+ \sum_{j =1}^{6} \left[ i f_j(t) \left|e_j \right> \left<e_{j+1} \right| + \text{H.c.} \right],
\nonumber %\label{eq:HamiltonianBenzene}
\end{eqnarray}
where $\left|e_j \right> $ denotes the excitation on site $j$ (defined modulo $6$),  $J_{j,j} = E_0$ is the on-site energy, $J_{j,j'} =\delta_{j,j'\pm1} J_0$ is the tunneling constant, and $f_j(t) =  f_{\Omega}\cos(\Omega t + 2\pi j/6 ) $ is the time-dependent tunneling strength with the driving amplitude  $f_{\Omega}$. The driving terms are motivated by the Peierls substitution describing a vectorial current-gauge-field coupling $\boldsymbol j \cdot \boldsymbol A(t)$~\cite{Bernevig2013} with a circularly rotating vector potential $\boldsymbol A(t)$.
The 
dipole transition operator $\hat V = \sum_{j=1}^{N}d_0 \left|e_j \right>  \left< g \right| $ excites the ground state  $\left|g \right> $  to the  single-excitation manifold, whose quasienergies are depicted in  Fig.~\ref{figSpectraBenzene}(b).
The stationary state is  $\rho_s(t) = \left|g\right>  \left< g\right|$ in agreement with Eq.~\eqref{eq:stationaryState}, i.e., a Floquet-Gibbs state for low  temperatures. 
A rotational symmetry is fulfilled for $N=6$ and $\hat R = \sum_{j=1}^{n} \left|e_{j+1}\right> \left< e_j \right|$.

In Fig.~\ref{figSpectraBenzene}(c), we depict the susceptibility  $\tilde \chi_0(\omega_{ p})$ of the benzene model. The resonances of the dark states defined by Eq.~\eqref{eq:symmetryProtectedDarkStates} are marked by dashed lines (optically invisible), and only two transitions, $\hat	V_{0 , 1} ^{(0)}\hat	V_{1 , 0} ^{(0)}$ and $\hat	V_{3 , 0} ^{(1)}\hat	V_{0 , 3} ^{(-1)}$, are visible. An aDS can be found for  $\hat	V_{0 , 1} ^{(0)}\hat	V_{1 , 0} ^{(0)}$ at 
$f_\Omega = 1.5 \Omega$.
As a consequence of the spDB in Eq.~\eqref{eq:floquetBandSelectionRule}, only Floquet bands $\tilde \chi_n(\omega_{ p})$ with  $n \mod  6 = 0$ appear. 

\textit{Parity symmetry.}
A dynamical parity symmetry is a specification of the dynamical rotational symmetry with $N=2$ and  a Hermitian operator $R^\dagger = R $ such that the spDS condition Eq.~\eqref{eq:symmetryProtectedDarkStates} and the spDB condition Eq.~\eqref{eq:floquetBandSelectionRule} are equally valid. The spDSs will be illustrated for the two-level system (TLS) in Eq.~\eqref{eq:Hamiltonian} along with the siT discussed below.

\begin{figure}[t]
	\includegraphics[width=0.9\linewidth]{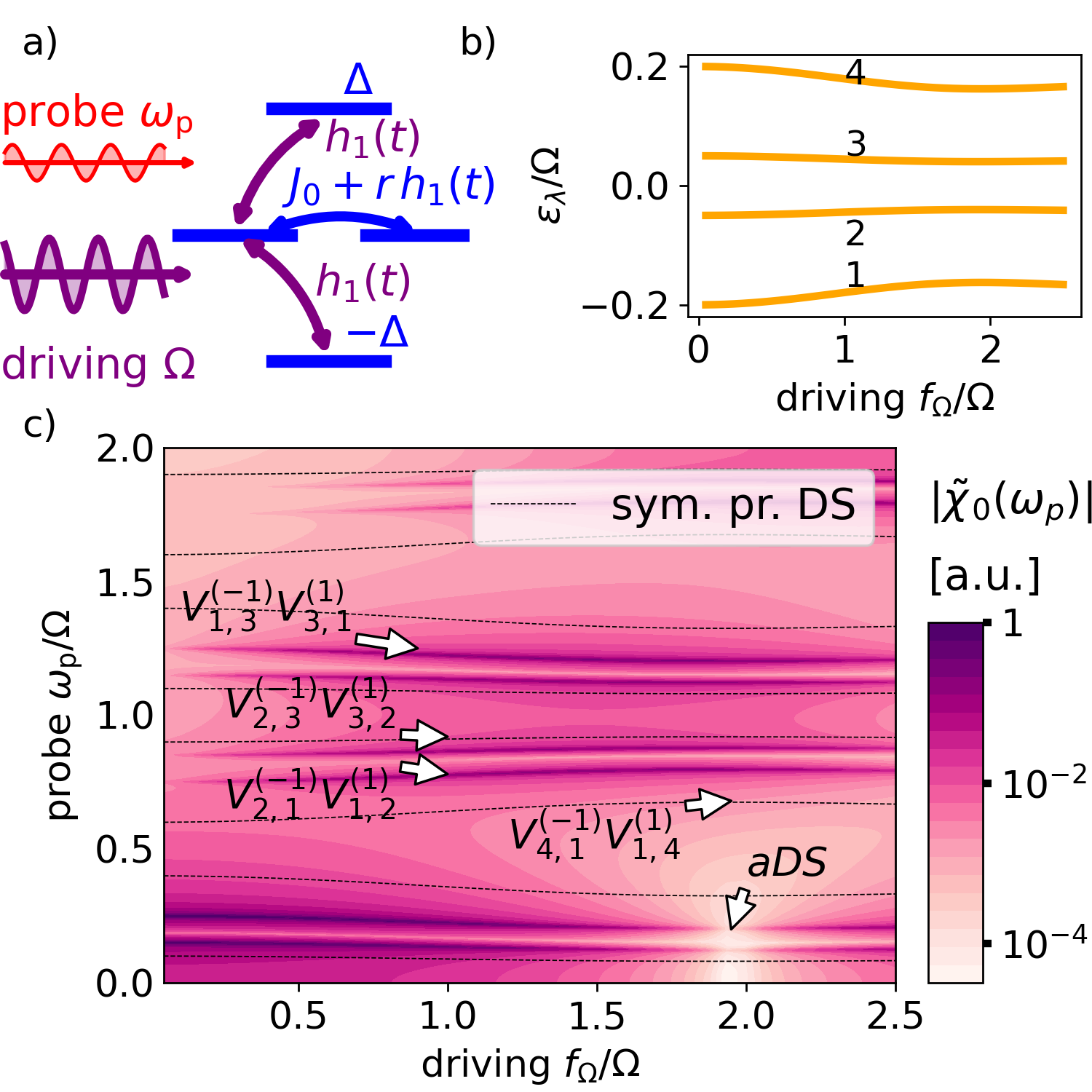}
	\caption{(a) Sketch of the dimer  model Eq.~\eqref{eq:hamiltionianFourStateModel2} with $h_1(t)  = f_{\Omega} \cos(\Omega t )  $.  (b) Quasienergy spectrum for  $J_0 /\Omega = 0.05$, $r  = 2 ,$ and $\Delta =0.2 \Omega$. (c) The  susceptibility $ \left| \tilde \chi_0(\omega_{ p})\right| $ is depicted as a color gradient. The spDSs  (marked by dashed lines) are generated by a particle-hole symmetry. }
	\label{figSpectraDimerModel}
\end{figure}%

\begin{figure}[t]
	\includegraphics[width=0.9\linewidth]{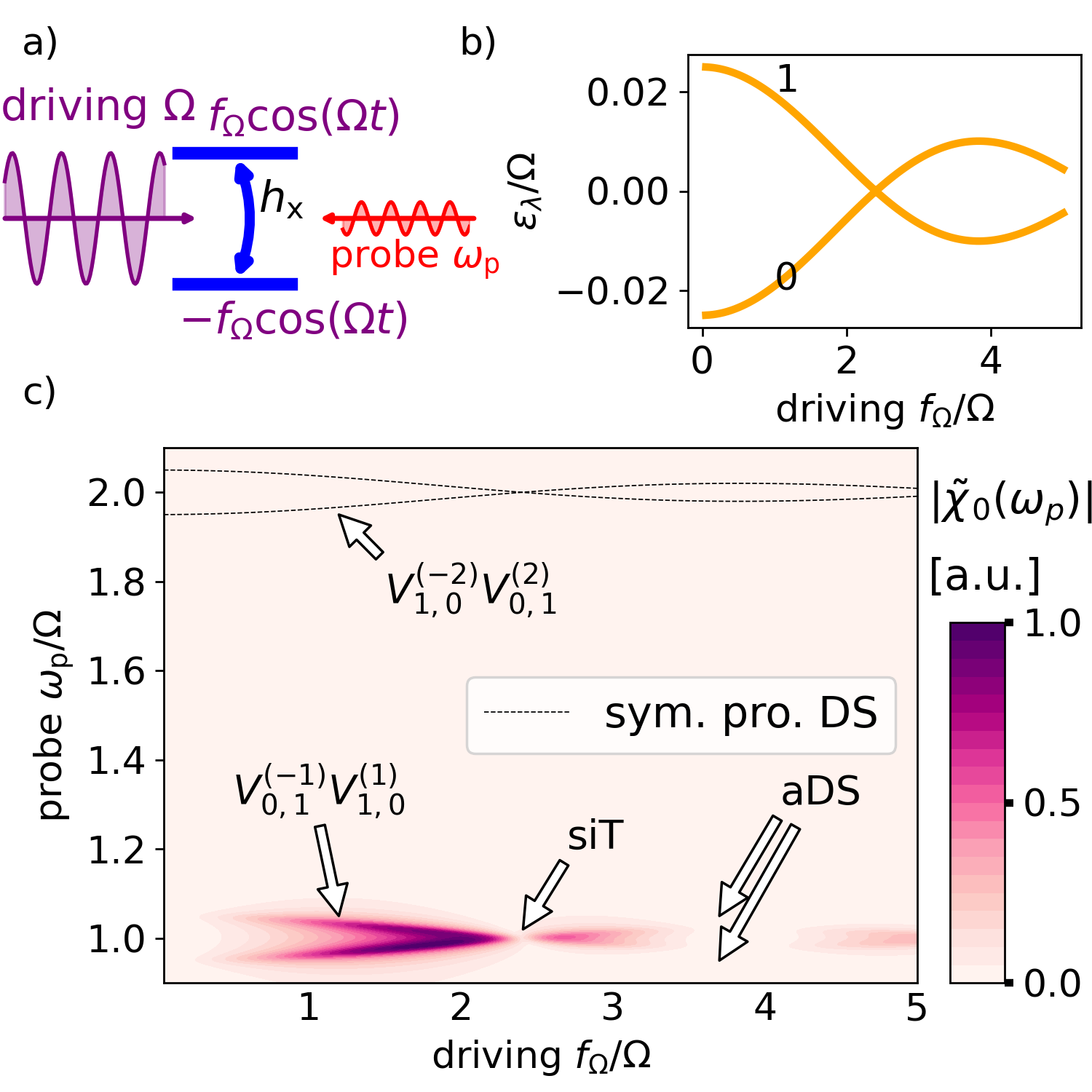}
	\caption{(a) Sketch of the ac-driven TLS.  (b) Quasienergy spectrum for $h_{ x} /\Omega = 0.05$. (c) The spectrum of the susceptibility $\tilde \chi_0(\omega_{ p})$ exhibits a siT and spDSs.  Here, $p_0= 0.6$ and $p_1 = 0.4$ in Eq.~\eqref{eq:stationaryState}  to highlight the siT. }
	\label{figSpectraSpinBoson}
\end{figure}%

\textit{Particle-hole symmetry.}
A particle-hole symmetry is defined for $-\alpha_S=\beta_S = 1 $, $t_S = t_P = \tau N_1/2N_2 $ with integers $N_1\in \left\lbrace  0,1\right\rbrace$, $N_2\geq 1$, and $\hat \Sigma= \hat   P\hat\kappa $ with a unitary operator $\hat P$ and the complex conjugation operator $\hat \kappa$, such that
$
	\hat P \hat  H^{*}(t+ t_{P}  ) \hat P =- \hat H(t) .
$
The particle-hole symmetry establishes a symmetry between the excitation and deexcitation processes and has its origin in fermionic systems, where adding and removing quasiparticles results in physically equivalent behaviors. Here we use the particle-hole symmetry in a general context.
 Using the particle-hole symmetry in Eq.~\eqref{eq:FloquetEquation}, we find that each Floquet state $\left|u_\mu (t)\right> $ with quasienergy $\epsilon_\mu$  has its symmetry related partner 
$
  \left|u_{\mu'} (t)\right>  = \pi^{(P)}_{\mu} \hat P \left|u_\mu (t+t_P)\right>^{*}\nonumber
$
with energy $ \epsilon_{\mu'} =- \epsilon_\mu $ and a gauge-dependent $\pi^{(P)}_{\mu}$. For $t_P = \tau/(2N_2)$,  the particle-hole symmetry gives rise to a rotational symmetry defined by $\hat R=\hat P \hat P $ and $t_R= \tau/N_2$ such that the dark state selection rules of the rotational symmetry apply. 
The particle-hole symmetry can give rise to a distinct dark state condition.
For  a dipole transition operator with  $\hat P^\dagger \hat  V^* \hat P = \alpha_V^{(P)}  \hat V $, $\alpha_V^{(P)}=\pm 1 $, $t_{P} = 0,\tau/2$, and $\hat P ^*\hat P= 1$, the particle-hole symmetry results in $V_{\mu ,\mu' }^{(m)}= \alpha_V^{(P)} e^{  i m \Omega t_{P}  }   V_{\mu ,\mu'  }^{(m)}$ for the symmetry-related states $\mu,\mu'$, so that
\begin{equation}
\hat	V_{\mu ,\mu'} ^{(m)} \propto 
\begin{cases}
0   &\text{if}\; \alpha_V^{(P)}e^{  i m \Omega t_{P}  }  =-1;\;  \mu ,\mu'  \text{  sym. rel.} \\ 
1  & \text{else} ,
\end{cases}
\label{eq:DarkStateCondtionParticleHole}
\end{equation}
as shown in detail in the Supplementary Material~\cite{supplementals}.
In contrast to Eq.~\eqref{eq:symmetryProtectedDarkStates}, where each transition can vanish for an appropriate $m$, only  transitions between symmetry-related states are affected by Eq.~\eqref{eq:DarkStateCondtionParticleHole}.

To illustrate Eq.~\eqref{eq:DarkStateCondtionParticleHole}, we use the dimer  model sketched in Fig.~\ref{figSpectraDimerModel}(a), 
with the Hamiltonian given by
\begin{eqnarray}
	H_{0}(t) =&&\Delta    \left( \hat A_{f,f} -\hat A_{g,g} \right) +  J_0 \hat A_{e_1,e_2}  \nonumber  \\
	&+& h_1(t) \left[ \hat A_{e_1,f} + \hat A_{g,e_1} + r \hat A_{e_1,e_2} \right]
	\label{eq:hamiltionianFourStateModel2},
\end{eqnarray}
 where $\hat A_{\alpha,\beta} \equiv \left|\alpha \right> \left < \beta\right|+ \text{H.c.}$, and $g,e_1,e_2$ and $f$ label the ground state, two single-excitation states, and  the double-excitation  state, respectively. $\Delta$ is the excitation gap,  $J_0$ is the tunneling constant, and $h_1(t) = f_{\Omega}  \cos(\Omega t)  $ is the driving field. The $r$ term   enhances higher-order dipole elements $V^{m\neq 0}_{\mu,\mu'}$. The particle-hole symmetry  is defined by $\hat P = \hat A_{g,f} + \hat A_{e_1,e_1}  - \hat   A_{e_2,e_2} $ and $t_P =0$. The quasienergy spectrum in Fig.~\ref{figSpectraDimerModel}(b) is symmetric with respect to $E =0$. The dipole transition operator is $\hat V= \hat A_{e_1,f} + \hat A_{g,e_1} $, such that $ \hat P^{\dagger} \hat V^{*} \hat P = - \hat V $. In Fig.~\ref{figSpectraDimerModel}(c), we depict the susceptibility in Eq.~\eqref{eq:dynamicSusceptibiltyFrequency}. According to the above considerations, the transitions between the particle-hole symmetry-related pairs vanish, i.e.,   $V_{1,4}^{(m)}=V_{4,1}^{(m)}=V_{2,3}^{(m)}=V_{3,2}^{(m)}= 0 $ for all $m$. These resonances are marked by dashed lines. The other transitions  not affected by the symmetry constrain remain visible in Fig.~\ref{figSpectraDimerModel}(c). 

\textit{Symmetry-induced transparency.}
The particle-hole symmetry can also give rise to a siT at the quasienergy crossing  $\epsilon_\mu = \epsilon_{\mu'} =0$ of the symmetry-related Floquet states $\mu,\mu'$. While a spDS is generated by a vanishing dipole element, $V_{\lambda,\mu }^{(n)}= 0$, the siT is generated by a destructive interference of two transitions with  $V_{\lambda,\mu }^{(n)}\neq 0$.
As shown in the Supplementary Material~\cite{supplementals} in detail, for two distinct particle-hole symmetries $\hat P_1 \neq \pm \hat P_2$, $\hat P_i^2 =\mathbbm 1 $ and $\left[\hat P_1, \hat P_2\right] =0$, the siT condition reads 
\begin{equation}
\tilde	\chi_n(m\Omega) \propto 
\begin{cases}
0   &\text{if}\; e^{i m \Omega (t_{P_1} - t_{P_2} )} = 1; \epsilon_\mu =\epsilon_{\mu'}=0 \\ 
1  & \text{else} ,
\end{cases}
\label{eq:transparencyCondition}
\end{equation}
where $t_{P_i} $ denote the reference times related to $\hat P_i$.

For illustration,  we consider an ac-driven TLS sketched in Fig.~\ref{figSpectraSpinBoson}(a) and described by the Hamiltonian 
\begin{align}
	\hat H_0(t)&= \frac{h_{\rm x} }{2}\hat \sigma_{\rm x} + \frac{f_{\Omega }}{2}  \cos\left(\Omega t \right) \hat \sigma_{\rm z} , 
	\label{eq:Hamiltonian}
\end{align}
where $\hat \sigma_x,\hat \sigma_{\rm z}$ are the Pauli matrices, $h_x$ is the tunneling amplitude, and $f_{\Omega}$ the driving strength. The TLS is weakly dissipative,   as in the  spin-boson model, such that it reaches the stationary state in Eq.~\eqref{eq:stationaryState}.   The dipole transition operator in Eq.~\eqref{eq:fullHamiltonian}  is $\hat V = \hat  \sigma_x$.  For $\hat R = \hat \sigma_{\rm x} $ and $ t_R =\tau/2$, the TLS exhibits a dynamical parity symmetry defined above, which gives rise to the coherent destruction of tunneling effect at an exact quasienergy crossing, depicted  in Fig.~\ref{figSpectraSpinBoson}(b) at $f_\Omega \approx 2.4\Omega$~\cite{Grossmann1991,Gong2009a}, and enables  the siT in the current context. Additionally, the TLS exhibits spDSs and spDBs according to Eq.~\eqref{eq:symmetryProtectedDarkStates} and Eq.~\eqref{eq:floquetBandSelectionRule} as $V_{\mu,\nu'  }^{(m)}=0$ for even $m$  because of the dynamical parity symmetry.

 For the TLS, a particle-hole symmetry is defined for $\hat P_1 = \hat \sigma_{\rm z} $ and $t_{P_1} = \tau/2$.  For $h_{\rm x} =0$, a second particle-hole symmetry is given for $\hat P_2 = \mathbbm 1  $ and $t_{P_2} = \tau/2$. As in this case $\epsilon_\mu =0$ and $ \hat P_i \hat \sigma_x^{*}  \hat P_i = (-1)^i \hat  \sigma_x  $,  siT with $\tilde	\chi_n(m\Omega) = 0$ appears according to Eq.~\eqref{eq:transparencyCondition}, and the response $\tilde \chi_{n}(\omega_{\rm p})$ is complete suppressed for all $n$.
In Fig.~\ref{figSpectraSpinBoson}(c), we consider $\tilde \chi_0(\omega_{ p})$ for a finite but small  $h_{\rm x}\ll \Omega$ such that the  quasienergy degeneracy is lifted except of the crossing, and  the particle-hole symmetry $\hat P_2$ is slightly broken.  As a consequence, the siT is not complete but scales as $\tilde	\chi_n(m\Omega)\propto h_x / \Omega $ at the crossing. 

\textit{Time-reversal and chiral symmetries.}
 A time-reversal symmetry (chiral symmetry) is defined by Eq.~\eqref{eq:symmetryRelationFloquetSpace} for $\alpha_{S}=-\beta_S=1$ ($\alpha_{S}= \beta_S=-1$), arbitrary $t_S$, and $\hat \Sigma= \hat  T \hat \kappa  $, ($\hat \Sigma= \hat  C$), where $\hat T $ ($\hat C $) is a unitary operator. 
As shown in the Supplementary Material~\cite{supplementals}, neither time-reversal symmetry nor chiral symmetry  alone implies spDSs.  However, the combination of time-reversal symmetry and chiral symmetry defines a particle-hole symmetry with $\hat P =  \hat C \hat T $, and $t_P = t_{T} - t_{C} $. When they further fulfill $t_{T} - t_{C } \in \left\lbrace   0,\tau/2 \right\rbrace $, $\hat C^*  \hat C=\mathbbm 1$, $\hat T^*  \hat T=\mathbbm 1$, and  $\left[\hat C, \hat T\right]= 0$, such that $\hat P^*\hat P =1$, spDSs  appear because of the particle-hole symmetry. In general, the presence of any two symmetries out of particle-hole symmetry, chiral symmetry, and time-reversal symmetry  implies  the existence of the third one. 

\textit{Conclusions.}
Using a unified conceptional framework based on Floquet response theory, we have predicted  selection rules in periodically driven quantum systems, namely  accidental dark states, symmetry-protected dark states,  symmetry-protected dark bands, and symmetry-induced transparency. The latter three effects are protected by symmetries such that variations of symmetry preserving parameters  do not destroy them. These symmetry-induced selection rules result from the destructive interference of a driven system synchronized to the periodic driving.
The different effects have been illustrated in three example systems fulfilling different symmetries, demonstrating the flexibility and generality of  our unified framework. The predicted selection rules are valid even for more complicated and realistic systems as long as the corresponding dynamical symmetries are fulfilled.

Our theoretical results are experimentally observable in systems that can reach the strong light-matter coupling regime such as cold-atom experiments~\cite{Yin2020} and
 superconducting circuits   \cite{Wang2019,Zha2020,Magazzu2018a,Chen2020}. For experiments with molecules, strong driving fields are necessary to generate high-order Floquet bands, but in cavity QED or  plasmonic fields, the strong driving interaction condition can be relaxed for  molecule ensembles interacting collectively with the light field~\cite{Herrera2016,Li2020a}.

\textit{Acknowledgements.} G. E. gratefully acknowledges financial support from the China Postdoc Science Foundation (Grant No. 2018M640054 ) and the Natural Science Foundation of China (Grant Nos. 11950410510  and  U1930402), J. C. acknowledges support from the NSF (Grant Nos. CHE 1800301 and  CHE1836913). The authors thank Tao  Wang for helpful discussions.

\bibliography{mybibliography}

\begin{thebibliography}{56}
\bibitem [{not()}]{noteRotationalSymmetry}%
  \BibitemOpen
  \href@noop {} {}\bibinfo {note} {One can construct a Hamiltonian fulfilling a
  time-rotational symmetry defined by $ \hat H_0(t) = \sum_j \sum_m f_j\left(t-
  m \frac{\tau}{N}\right)\hat A_{j}^{(m)},$ where $\hat A_{j}^{(m)} = \hat R^m
  \hat A_{j}^{(0)}\hat R^m $ for arbitrary Hermitian operators $\hat
  A_{j}^{(0)} $, and the arbitrary functions $ f_j\left(t+\tau\right) =
  f_j\left(t\right)$ shall be $\tau$ periodic.}\BibitemShut {Stop}%
\end{thebibliography}

\begin{widetext}
	
	\begin{center}
		\Huge
		\textbf{
		Supplementary Material}
	\end{center}
	
	\section{Floquet equation and dynamical symmetries}

	In this supplemental information, we provide details to establish the symmetry-protected dark state conditions, which are sketched in the main text. The starting point is the Floquet equation 
	\begin{equation}
	\left[  \hat H_0(t)-i\frac{d}{dt} \right] \left| u_\mu (t)\right>   =   \epsilon_\mu \left| u_\mu (t)\right> ,
	\label{eq:FloquetEquationSI}
	\end{equation}
	and the time-spatial symmetry relation for a Hamiltonian in the Floquet space
	\begin{equation}
	\hat  \Sigma  \left[ \hat H_0(t_s+ \beta_S t)-i\frac{d}{dt}  \right] \hat \Sigma ^{-1}  =   \alpha_S \left[   \hat H_0(t)-i\frac{d}{dt}  \right],
	\label{eq:symmetryRelationFloquetSpaceSI}
	\end{equation}
	where the symmetry is specified by the spatial operator $\hat \Sigma$, the time-shift  $t_S$, $\alpha_S =\pm 1$ and $\beta_S =\pm 1$.

	\section{Dynamical Dipole elements }
	
	The susceptibility, introduced in the main text,
	\begin{equation}
	\tilde \chi_{n}(\omega ) = i \lambda^2 \sum_{\nu,\mu,m} \frac{ V_{\nu,\mu }^{(-n-m)}  V_{\mu ,\nu}^{(m)}\left(p_\nu -p_\mu \right) }{ \epsilon_\mu -\epsilon_\nu +m\Omega  - \omega - i \gamma_{\nu,\mu }^{(m)}    } 
	\label{eq:dynamicSusceptibiltyFrequencySI},
	\end{equation}
	is expressed in terms of the dynamical dipole elements
	\begin{equation}
	V_{\mu ,\nu }^{(n)} =  \frac{1}{\tau} \int_{0}^{\tau}\left<u_{\mu}\right|  \hat V (t) \left|  u_{\nu}\right>e^{ - i n \Omega t} dt,
	\label{eq:dynamicalDipoleElementsSI}
	\end{equation}
	which fulfill
	\begin{eqnarray}
	V_{\mu ,\nu }^{(n)} &=& \frac{1}{\tau} \int_{0}^{\tau}  \left< u_{\mu} (t)  \right| \hat V    \left|u_{\nu}  (t)\right> e^{ - i n \Omega t}dt , \nonumber \\
	&=& \frac{1}{\tau} \int_{0}^{\tau} \left(  \left< u_{\nu} (t)  \right| \hat V   \left|u_\mu  ( t)\right> \right)^{*} e^{ - i n \Omega t}dt , \nonumber \\
	&=& \left( \frac{1}{\tau} \int_{0}^{\tau}  \left< u_{\nu} (t)  \right| \hat V   \left|u_\mu  ( t)\right>  e^{  i n \Omega t}dt, \right)\nonumber  \\
	&=&  \left[ V_{\nu ,\mu }^{(-n)} \right]^*.
	\label{eq:generalSymmetryCondtionSI}
	\end{eqnarray}
	This relation will be  used to prove the symmetry-protected dark state condition based on a dynamical particle-hole symmetry in Sec.~\ref{sec:particleHoleDarkStateSI}.

	\section{Rotational symmetry}
	
	\label{sec:darkStatesParityProofSI}
	
	Here we present the derivation of the dark state condition imposed by a dynamical rotational symmetry. For a unitary  operator $\hat \Sigma =\hat R$, $t_S = t_P = \tau/N $ with a positive integer $N$, and $\alpha_S=\beta_S=1$, the general dynamical symmetry condition Eq.~\eqref{eq:symmetryRelationFloquetSpaceSI} specifies  to a rotational symmetry
	\begin{equation}
	\hat R  \hat 	H\left(t+\frac \tau N\right)  \hat R^\dagger  = \hat H( t ),
	\label{eq:dynRotSymmetrySI}
	\end{equation}
	Applying Eq.~\eqref{eq:dynRotSymmetrySI} to Eq.~\eqref{eq:FloquetEquationSI},  we find
	\begin{equation}
	\left[  \hat H_0(t)-i\frac{d}{dt}  \right] \hat R \left| u_\mu (t+t_R )\right>   =   \epsilon_\mu \hat R  \left| u_\mu (t+t_R  )\right> .
	\label{eq:FloqueeEquationRotationSI}
	\end{equation}
	This implies that every Floquet state is also an eigenstate of the rotational symmetry operator, such that
	\begin{equation}
	\hat R \; \left| u_\mu \left(t+t_R   \right) \right> = \pi_\mu^{(R)} \left| u_\mu (t) \right>,
	\label{eq:parityEigenvalueEquationIIsI}
	\end{equation}
	with eigenvalues $\pi_\mu^{(R)} = e^{i 2\pi m_\mu /N  }$ and integers $m_\mu =\left\lbrace 0,N-1 \right\rbrace  $.   We require that the transition dipole operator obeys $\hat R^{\dagger} \hat V \hat R = \alpha_V^{(R)} \hat V$ with $\alpha_V^{(R)}=\pm 1$.
	Combining with Eq.~\eqref{eq:FloquetEquationSI}, the dynamical dipole element fulfills
	\begin{eqnarray}
	V_{\nu,\mu }^{(n)} &=& \frac{1}{\tau} \int_{0}^{\tau}  \left< u_\nu (t)  \right| \hat V    \left|u_\mu  (t)\right> e^{ - i n \Omega t}dt , \nonumber \\
	&=& \frac{1}{\tau} \sum_{m=0}^{N-1} \int_{m\tau/N }^{(m+1)\tau/N}\left< u_\nu (t)  \right| \hat V    \left|u_\mu  (t)\right> e^{ - i n \Omega} t dt , \nonumber \\
	&=& \frac{1}{\tau} \sum_{m=0}^{N-1} \int_{0 }^{\tau/N}\left< u_\nu \left( t+m\frac \tau N  \right)  \right| \hat V    \left|u_\mu  \left( t+m\frac \tau N  \right)\right> e^{ - i n \Omega \left( t+m \frac \tau N  \right)}dt , \nonumber \\
	&=& \frac{1}{\tau} \sum_{m=0}^{N-1} e^{ - i 2\pi m n  / N  }\int_{0 }^{\tau/N}   \left[ \left( \pi_\mu^{(R)}\right)^*    \pi_\nu ^{(R)} \right] ^m\left< u_\nu \left( t \right)  \right| \hat R^{\dagger m} \hat V  \hat R^{m}    \left|u_\mu  \left( t\right)\right> e^{ - i n \Omega t }dt , \nonumber \\
	&=&  \frac {V_{\nu,\mu }^{(n)} }N \sum_{m=0}^{N-1} e^{ - i 2\pi n m  / N  }    \left( \alpha_V^{(R)}  \right)^{m}   e^{ i m\left( m_\nu -m_\mu\right) 2\pi /N}   \nonumber .
	\end{eqnarray}
	Recalling that $\alpha_V^{(R)} =\pm 1$, the last line establishes  the dark state condition for a dynamical rotational symmetry 
	\begin{equation}
	V_{\nu ,\mu} ^{(n)} \propto 
	\begin{cases}
	1   &\text{if}\; e^{i \frac{2\pi}{N}\left( m_\mu-m_\nu+ n \right) }\alpha_V^{(R)}  =1, \\ 
	0  & \text{else} ,
	\label{eq:symmetryProtectedDarkStatesSI}
	\end{cases}
	\end{equation}
	which is presented in the main text.

	\section{Pariticle-hole symmetry}

	\label{sec:particleHoleDarkStateSI}
	
	Here, we provide details to establish the dark state condition induced by a dynamical particle-hole symmetry 
	\begin{equation}
	\hat P  \hat	H^{*}(t_{P} +t)  \hat P^\dagger  = -\hat  H( t ),
	\label{eq:defintionParticleHoleSymmetrySI}
	\end{equation}
	where $t_{P}$ assumes the values $t_{P} =N_1\tau/2N_2$ for integers $N_1 =0,1$ and $N_2\geq 1$ and depends on the specific system. We obtain Eq.~\eqref{eq:defintionParticleHoleSymmetrySI} from the general definition of the dynamical symmetries Eq.~\eqref{eq:symmetryRelationFloquetSpaceSI} for $\hat \Sigma =  \hat P \hat \kappa$, with the complex conjugation operator $\hat \kappa$ and the unitary operator $\hat P$, $\alpha_S= -\beta_S=-1$ and $t_S= t_P$.
	Applying the definition in Eq.~\eqref{eq:defintionParticleHoleSymmetrySI} to the Floquet equation Eq.~\eqref{eq:FloquetEquationSI}, we find
	\begin{equation}
	\left[ \hat H_0(t)-i\frac{d}{dt}  \right] \hat P \left| u_\mu (t_{P} + t )\right>^{*}   =  - \epsilon_\mu \hat P  \left| u_\mu (t_{P} + t )\right>^{*} ,
	\label{eq:FloquetEquationParticleHoleSI}
	\end{equation}
	which indicates  that for every Floquet state $ \left| u_\mu (t )\right> $ with quasienergy $\epsilon_\mu$, there is a symmetry-related partner
	\begin{equation}
	\left| u_{\mu'} ( t)\right> = \pi_{\mu }^{(P)}   \hat P   \left| u_{\mu} (t_{P} + t)\right>^{*}
	\label{eq:particleHoleFloquetStateRelationSI}
	\end{equation}
	with quasienergy $\epsilon_{\mu'} =-  \epsilon_\mu$, and a gauge-dependent phase factor $\pi_{\mu }^{(P)}$. The phase factor cannot be removed by a simple gauge transformation as the two  Floquet states $\mu$ and $\mu'$ are coupled. However, we can apply Eq.~\eqref{eq:particleHoleFloquetStateRelationSI} twice and obtain
	\begin{equation}
	\left| u_{\mu} ( t)\right> = \pi_{\mu }^{(P)} \pi_{\mu' }^{(P)*} \hat P  \hat P^{*}   \left| u_{\mu} (2 t_{P} + t)\right>.
	\label{eq:particleHoleRotatinalSymmetrySI}
	\end{equation}
	In general, $\pi_{\mu }^{(P)} \neq \pi_{\mu' }^{(P)} $, and we cannot find a gauge transformation such that $\pi_{\mu }^{(P)} \pi_{\mu' }^{(P)*} =1$. 
	Only if $\hat P  \hat P^{*} =\mathbbm 1$ and $t_{P} \in \left\lbrace 0 ,\tau/2 \right\rbrace$,  we have $\pi_{\mu' }^{(P)*} \pi_{\mu }^{(P)}=1 $, which will be used in the evaluation of the dynamical dipole elements. Furthermore, we require $\hat P^{\dagger}  \hat V \hat P  = \alpha_V^{(P)} \hat V^{*} $ with $\alpha_V^{(P)} = \pm 1$.
	Using Eq.~\eqref{eq:particleHoleFloquetStateRelationSI} and Eq.~\eqref{eq:particleHoleRotatinalSymmetrySI},  we  find that the dynamical dipole element obeys
	\begin{eqnarray}
	V_{\mu ,\mu' }^{(n)} &=& \frac{1}{\tau} \int_{0}^{\tau}  \left< u_{\mu} (t)  \right| \hat V    \left|u_{\mu'}  (t)\right> e^{ - i n \Omega t}dt  \nonumber \\
	&=& \frac{1}{\tau} \int_{0}^{\tau} \pi_{\mu' }^{(P)*} \pi_{\mu }^{(P)}  \left< u_{\mu'} ( t + t_{P})  \right|^{*}  \hat P^{\dagger} \hat V \hat P   \left|u_\mu  ( t+t_{P})\right>^{*} e^{ - i n \Omega t}dt  \nonumber \\
	&=& \frac{1}{\tau} \int_{0}^{\tau} \alpha_V^{(P)} \left< u_{\mu'} (t)  \right|^{*} \hat V^{*}   \left|u_\mu  (t)\right>^{*}  e^{ - i n \Omega ( t-t_{P})}dt  \nonumber \\
	&=& \frac{1}{\tau}   \alpha_V^{(P)} e^{ - i n \Omega t_{P}  } dt \int_{0}^{\tau} \left(  \left< u_{\mu'} (t)  \right| \hat V   \left|u_{\mu}  (t)\right> \right)^{*} e^{  i n \Omega t}dt  \nonumber \\
	&=& \alpha_V^{(P)} e^{ - i n \Omega t_{P}  }   V_{\mu',\mu }^{(-n)*}
	= \alpha_V^{(P)}  e^{ - i n \Omega t_{P}  }   V_{\mu,\mu' }^{(n)},
	\label{eq:dynamicalDipolElementsSymmetrySI}
	\end{eqnarray}
	where we have used Eq.~\eqref{eq:generalSymmetryCondtionSI} in the last line. From line two to line three we have used the $\tau$-periodicity of the integrand. This proves the particle-hole symmetry  induced condition for dark states
	\begin{equation}
	\hat	V_{\mu ,\mu'} ^{(m)} \propto 
	\begin{cases}
	0   &\text{if}\; \alpha_V^{(P)}e^{ - i n \Omega t_{P}  } =-1,\quad  \mu ,\mu'  \text{  sym. rel.} \\ 
	1  & \text{else} ,
	\end{cases}
	\end{equation}
	
	which is presented  in the main text.

\section{Chiral symmetry}
	\label{sec:ChrialSymmetrySI}
	
	Here we derive a constrain for the dynamical dipole elements under a dynamical chiral symmetry, and explain why the chiral symmetry on its own does not imply a symmetry-protected dark state condition.  For a unitary $\hat \Sigma =\hat C$, $\alpha_S=\beta_S=-1$
	and an arbitrary  $t_S = t_C$,  the general dynamical symmetry relation Eq.~\eqref{eq:symmetryRelationFloquetSpaceSI} defines a chiral symmetry
	\begin{equation}
	\hat C \hat	H(t_{C} - t)  \hat C^\dagger  = - \hat  H(t ).
	\end{equation}
	Applying this definition to the Floquet equation Eq.~\eqref{eq:FloquetEquationSI}, we find
	\begin{equation}
	\left[ \hat H_0(t)   -i\frac{d}{dt}  \right] \hat C \left| u_\mu (t_{C} - t)\right>   =  - \epsilon_\mu \hat C \left| u_\mu (t_{C} - t )\right> .
	\label{eq:FloquetEquationChiralSI}
	\end{equation}
	Similar to the particle-hole symmetry,  Eq.~\eqref{eq:FloquetEquationChiralSI} implies  that for every Floquet state $ \left| u_\mu (t )\right> $ with quasienergy $\epsilon_\mu$, there is a symmetry-related partner
	\begin{equation}
	\left| u_{\mu'} (t) \right> = \pi_{\mu }^{(C)}   \hat C   \left| u_{\mu} (t_{C}-  t)\right>,
	\label{eq:FloquetStatesChiralSymmetrySI}
	\end{equation}
	with quasienergy $\epsilon_{\mu'} =-\epsilon_\mu$ and a gauge-dependent phase factor $\pi_{\mu }^{(C)}$. The $\pi_{\mu }^{(C)}, \pi_{\mu '}^{(C)}$ can not be arbitrarily changed by a gauge transformation. Chiral symmetry is not sufficient to determine a constrain for the dynamical dipole elements. To see this we conjugate Eq.~\eqref{eq:FloquetStatesChiralSymmetrySI} and insert it into itself, and obtain 
	\begin{equation}
	\left| u_{\mu} (t) \right> =  \pi_{\mu }^{(C)}  \pi_{\mu' }^{(C)*}  \hat C \hat C^{*}  \left| u_{\mu} ( t)\right>^{*}.
	\end{equation}
	When requiring  $\hat C^{*} \hat C = \mathbbm 1$, we can find a gauge transformation, for which $\pi_{\mu }^{(C)}  \pi_{\mu' }^{(C)*}=1$. Using this and $\hat C^\dagger V \hat C = \alpha_V^{(C)} \hat V$ with $\alpha_V^{(C)} = \pm 1$ to evaluate the dynamical dipole elements, we find
	\begin{eqnarray}
	V_{\mu ,\mu' }^{(n)} &=& \frac{1}{\tau} \int_{0}^{\tau}  \left< u_{\mu} (t)  \right| \hat V    \left|u_{\mu'}  (t)\right> e^{ - i n \Omega t}dt  \nonumber \\
	&=& \frac{1}{\tau} \int_{0}^{\tau} \pi_{\mu' }^{(C)*} \pi{\mu' }^{(C)}  \left< u_{\mu'} ( t_{C}-  t )  \right|  \hat C^{\dagger} \hat V \hat C  \left|u_\mu  ( t_{C}-  t)\right> e^{ - i n \Omega t}dt  \nonumber \\
	&=& \frac{1}{\tau} \int_{0}^{\tau} \alpha_V^{(C)} \left< u_{\mu'} (t)  \right| \hat V   \left|u_\mu  (t)\right>  e^{ - i n \Omega (  t_{C}-  t )}dt  \nonumber \\
	&=& e^{ - i n \Omega   t_{C}}  \frac{1}{\tau}  \alpha_V^{(C)} \int_{0}^{\tau}  \left< u_{\mu'} (t)  \right| \hat V   \left|u_{\mu}  (t)\right>  e^{  i n \Omega t}dt  \nonumber \\
	&=& e^{ - i n \Omega   t_{C}}  \alpha_V^{(C)}    V_{\mu',\mu }^{(-n)}  \nonumber  \\
	&=& e^{ - i n \Omega   t_{C}}  \alpha_V^{(C)}    V_{\mu,\mu' }^{(n)*}  ,
	\label{eq:dynamicalDipolElementsChiralSymmetrySI}
	\end{eqnarray}
	where we have used Eq.~\eqref{eq:generalSymmetryCondtionSI} in the last line, and the $\tau$ periodicity of the integrand in line three.
	This relation alone does not imply a dark-state condition for  the dynamical dipole elements. If $ V_{\mu,\mu' }^{(n)*} $ was real (e.g., because of another symmetry relation), Eq.~\eqref{eq:dynamicalDipolElementsChiralSymmetrySI} would indeed imply a dark state condition, but chiral symmetry alone is not sufficient to guarantee this.

	\section{Time-reversal symmetry}
	\label{sec:TimreversalSymmetrySI}
	
	Similar to  the chiral symmetry, the time-reversal symmetry alone does not impose a dark state condition. A time-reversal symmetry in  Eq.~\eqref{eq:symmetryRelationFloquetSpaceSI} is defined for  $\hat \Sigma = \hat T \hat \kappa$, with a unitary operator $\hat  T $, the complex conjugation operator $\hat \kappa$,  $\alpha_S=-\beta_S=1$,
	and an arbitrary $t_S =t_T $ , so that
	\begin{equation}
	\hat  T \hat	H^{*}(t_{ T} - t)  \hat  T^\dagger  =  \hat  H( t).
	\end{equation}
	Applying this definition to the Floquet equation, we find
	\begin{equation}
	\left[  \hat H_0(t)-i\frac{d}{dt} \right] \hat T\left| u_\mu (t_{T} - t)\right>^{*}   =   \epsilon_\mu \hat T \left| u_\mu (t_{T} - t )\right>^{*} ,
	\label{eq:FloquetEquationTRSI}
	\end{equation}
	Thus, all Floquet states fulfill
	\begin{equation}
	\left| u_{\mu} (t) \right> = \pi_{\mu }^{(T)}   \hat T   \left| u_{\mu} (t_{T}-  t)\right>^{*} 
	\label{eq:FloquetStatesTRSymmetrySI}
	\end{equation}
	with a phase factor $\pi_{\mu }^{( T)}$, which can be transformed away by a gauge transformation. Using this to evaluate the dipole elements, we find
	\begin{eqnarray}
	V_{\mu ,\nu }^{(n)} &=& \frac{1}{\tau} \int_{0}^{\tau}  \left< u_{\mu} (t)  \right| \hat V    \left|u_{\nu }  (t)\right> e^{ - i n \Omega t}dt , \nonumber \\
	&=& \frac{1}{\tau} \int_{0}^{\tau}   \left< u_{\mu } ( t_{T}-  t )  \right|^{*}  \hat  T^{\dagger} \hat V \hat  T  \left|u_\nu  ( t_{T}-  t)\right>^{*} e^{ - i n \Omega t}dt , \nonumber \\
	&=& \frac{1}{\tau} \int_{0}^{\tau} \alpha_V^{( T)} \left< u_{\mu} (t)  \right|^{*} \hat V^{*}   \left|u_\nu  (t)\right>^{*}  e^{ - i n \Omega (  t_{ T}-  t )}dt , \nonumber \\
	&=& e^{ - i n \Omega   t_{ T}} \alpha_V^{(T)}  \left(   \frac{1}{\tau}  \int_{0}^{\tau}  \left< u_{\mu } (t)  \right| \hat V   \left|u_{\nu }  (t)\right>  e^{-  i n \Omega t}dt  \right)^{*} , \nonumber \\
	&=& e^{ - i n \Omega   t_{ T}}  \alpha_V^{( T)}    V_{\mu,\nu }^{(n)*}  ,\nonumber  \\
	\label{eq:dynamicalDipolElementsTRSymmetrySI}
	\end{eqnarray}
	Similar to the chiral symmetry, this does not establish a dark state condition.
	
	\section{Combined chiral symmetry and time reversal symmetry}
	
	Importantly, the chiral-symmetry relation Eq.~\eqref{eq:dynamicalDipolElementsChiralSymmetrySI} and the time-reversal symmetry relation Eq.~\eqref{eq:dynamicalDipolElementsTRSymmetrySI} together do  provide a sufficient condition for a symmetry-protected dark state. In the derivations in Sec.~\ref{sec:ChrialSymmetrySI}, and Sec.~\ref{sec:TimreversalSymmetrySI}  we could choose the gauge fields independently, such that $\pi_{\mu}^{(C)}\pi_{\mu}^{(C)*}=1 $ and $\pi_{\mu}^{(T)}\pi_{\mu}^{(T)*}=1 $ . Yet, these gauges conditions can not be simultaneous fulfilled in general.
	However, the combination of a time-reversal symmetry and a chiral symmetry defines a particle-hole symmetry with $\hat P  =  \hat C\hat T $ and $t_{P} =  t_{T } -t_{C} $. When we additionally require that $\hat C \hat C ^{*}=\mathbbm 1$, $\hat T \hat T  ^{*}=\mathbbm 1$, $\left[\hat C, \hat T  \right] = 0$, such that $\hat P ^*\hat P=\mathbbm 1$, and $t_{T } - t_{C} =0,\tau/2 $, all requirements for symmetry-protected dark states in Sec.~\ref{sec:particleHoleDarkStateSI} are fulfilled.

	\section{Symmetry-induced transparency}

	\begin{figure*}[t]
		\includegraphics[width=0.9\linewidth]{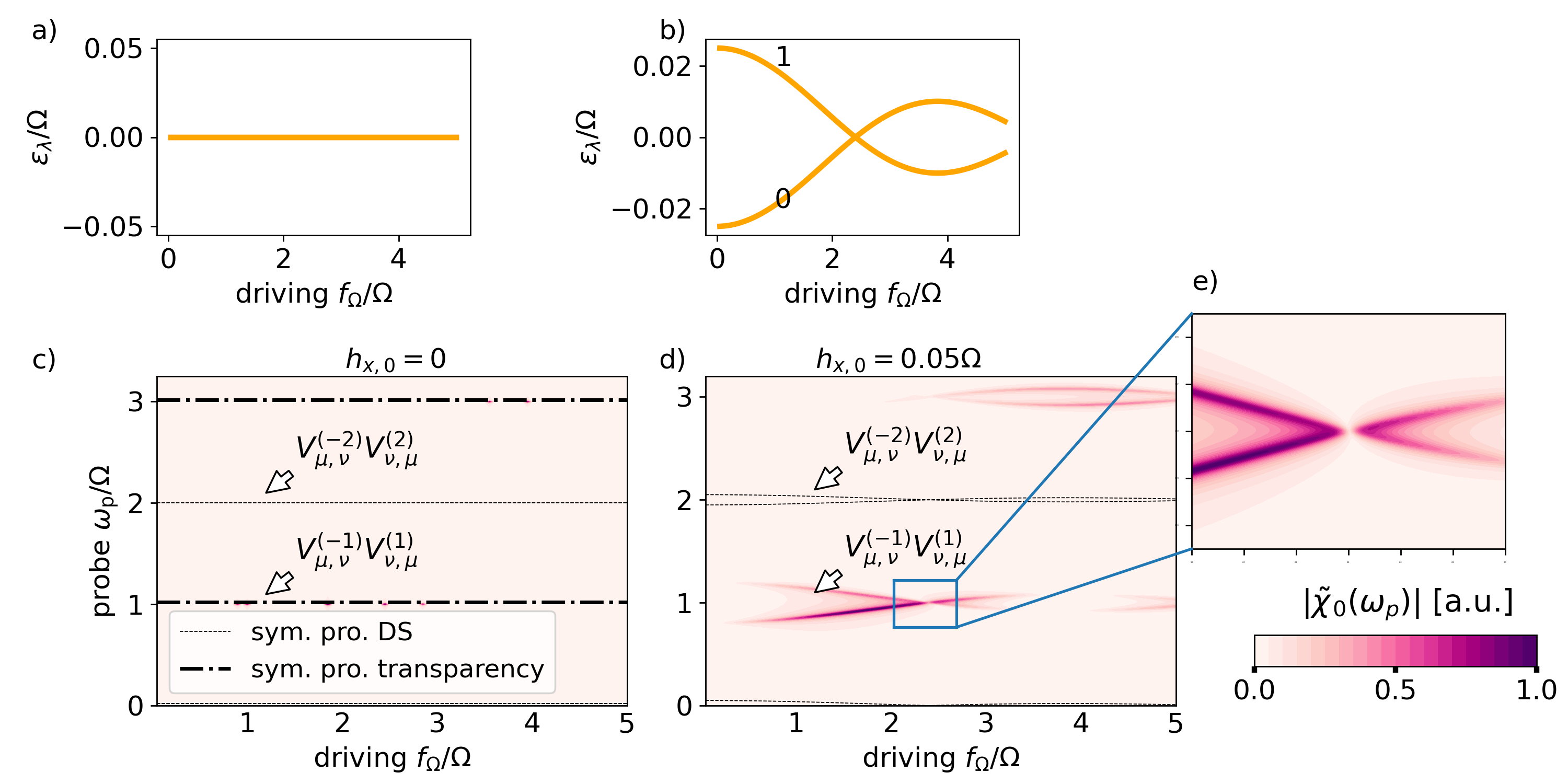}
		\caption{(a) and (b) depict the quasienergies  for $h_{\rm x} /\Omega = 0$ and $h_{\rm x} /\Omega = 0.05$, respectively. (c) and (d) show  the susceptibility $\tilde \chi_0(\omega_{\rm p})$ depicted by a color gradient for $h_{\rm x} /\Omega = 0$ and $h_{\rm x} /\Omega = 0.05$, respectively. Dynamical parity spDS are marked  by dashed lines, while transitions suppressed by the symmetry-protected transparency are marked by dashed-dotted lines. In (d), one particle-hole symmetry is slightly broken. Consequently, the symmetry-protected transparency becomes a symmetry-induced transparency (siT) appearing at a quasienergy crossing, which is magnified in panel (e).}
		\label{figSpectraSpinBosonSI}
	\end{figure*}%

	The symmetry-protected dark states and symmetry-protected dark bands introduced before are defined by  vanishing dynamical dipole elements in Eq.~\eqref{eq:dynamicalDipoleElementsSI}. Here we want to introduce a distinct transparency effect, which is also protected by dynamical symmetries. In order to distinguish it from the symmetry-protected dark states and symmetry-protected dark bands, we coin it symmetry-protected transparency. We also explain a weaker form of it, i.e., symmetry-induced transparency. 
	
	While a symmetry-protected dark state is generated by a vanishing dipole element, i.e., $V^{(m)}_{\mu,\nu}=0$, the symmetry-protected transparency is generated by a destructive interference of two transitions with non-vanishing dipole elements. More precisely, the transparency effect appears when for a given resonance $\omega\approx m\Omega$ the corresponding two terms in the susceptibility in Eq.~\eqref{eq:dynamicSusceptibiltyFrequencySI} cancel each other, which requires
	\begin{equation}
	V^{(-n-m)}_{\mu',\mu} V^{(m)}_{\mu,\mu'}  = V^{(-n-m)}_{\mu,\mu'} V^{(m)}_{\mu',\mu},
	\label{eq:transparencyRequirementNSI}
	\end{equation}
	which will be investigated in the following. 
	
	The transparency effect appears as a consequence of two distinct particle-hole symmetries, if two symmetry-related resonances corresponding to the quasienergies $\epsilon_{\mu}=-\epsilon_{\mu'}$ become degenerate, i.e., $\epsilon_{\mu'} =\epsilon_{\mu} = 0 $. 
	We denote the two particle-hole symmetries with $\hat P_1 = \hat P_1 ^{\dagger}$, $\hat P_2=\hat P_2^{\dagger}$ with corresponding time shifts $t_{P_1},t_{P_2} $, and require $\left[\hat P_1 ,\hat P_2 \right]=0$ as well as $\hat P_1 \neq \pm \hat P_2 $ (i.e., they are distinct).
	The Floquet states $\mu$ with $\epsilon_{\mu} =0$  form a subspace of dimension two. As $\hat P_1$ and $\hat P_2$ commute, and $\hat P_i = \hat P_i ^{\dagger}$, there is a basis of this subspace  such that
	\begin{eqnarray}
	\hat P_1 \left| v_\mu ( t_{P_1} +t )\right>^{*}	&=&   \left| v_\mu ( t )\right>, \nonumber \\
	\hat P_1 \left| v_{\mu'} (t_{P_1} + t )\right>^{*}	&=&   \left| v_{\mu'} ( t )\right>, \nonumber  \\ 
	\hat P_2 \left| v_\mu ( t_{P_2} +t )\right>^{*}	&=&   \left| v_\mu ( t )\right>,\nonumber  \\
	\hat P_2 \left| v_{\mu'} ( t_{P_2} +t )\right>^{*}	&=&  - \left| v_{\mu'} ( t )\right>. %
	\label{eq:phsBasisSI}
	\end{eqnarray}
	In this basis, there are no pairs of symmetry-related partners as introduced in Sec.~\ref{sec:particleHoleDarkStateSI}. From  Eq.~\eqref{eq:phsBasisSI} we obtain a basis with a symmetry-related pair by
	\begin{eqnarray}
	\left| u_\mu ( t )\right>	&=&  \frac 1{\sqrt{2}} \left(  \left| v_\mu (t )\right>  +  \left| v_{\mu'} (t )\right> \right), \nonumber \\
	\left| u_{\mu'} ( t )\right>	&=&  \frac 1{\sqrt{2}} \left(  \left| v_{\mu} ( t )\right> -  \left| v_{\mu'} (t )\right> \right),
	\end{eqnarray}
	which are related by the symmetry operations as
	\begin{eqnarray}
	\hat P_1 \left| u_{\mu'} ( t_{P_1} +t )\right>^{*}	&=&   \left| u_\mu (t )\right>, \nonumber  \\
	\hat P_1 \left| u_{\mu} (t_{P_1} + t )\right>^{*}	&=&   \left| u_{\mu'} ( t )\right>, \nonumber  \\ 
	\hat P_2 \left| u_\mu (t_{P_2} + t )\right>^{*}	&=&   \left| u_{\mu}( t )\right>,\nonumber  \\
	\hat P_2 \left| u_{\mu'} ( t_{P_2} +t )\right>^{*}	&=&   \left| u_{\mu'} ( t )\right>. %
	\end{eqnarray}
	Thus, we have established symmetry-related pairs  with respect to $\hat P_1$. Please note that the gauges phases are fixed by the procedure, such that we do not consider  them in the following calculations.
	Using the symmetry $\hat P_2$ to evaluate the dynamical dipole elements, we find
	\begin{eqnarray}
	V_{\mu ,\mu' }^{(n)} &=& \frac{1}{\tau} \int_{0}^{\tau}  \left< u_{\mu} (t)  \right| \hat V    \left|u_{\mu'}  (t)\right> e^{ - i n \Omega t}dt \nonumber \\
	&=& \frac{1}{\tau} \int_{0}^{\tau}   \left< u_{\mu} (t_{P_2} + t  )  \right|^{*}  \hat P_2^{\dagger} \hat V \hat P_2   \left|u_{\mu'}  ( t_{P_2} + t  )\right>^{*} e^{ - i n \Omega t}dt  \nonumber \\
	&=& \frac{1}{\tau} \int_{0}^{\tau} \alpha_V^{(P)} \left< u_{\mu} (t)  \right|^{*} \hat V^{*}   \left|u_{\mu'}  (t)\right>^{*}  e^{ - i n \Omega ( t_{P_1} + t )}dt  \nonumber \\
	&=&  \alpha_V^{(P_1)}  e^{-in\Omega  t_{P_1}   }  \left(  \frac{1}{\tau}  \int_{0}^{\tau}  \left< u_{\mu} (t)  \right| \hat V   \left|u_{\mu'}  (t)\right>  e^{  i n \Omega t}dt \right)^{*}  \nonumber \\
	&=& \alpha_V^{(P_1)}   e^{-in\Omega  t_{P_1}   } V_{\mu,\mu' }^{(-n)*} \nonumber.  \\
	\label{eq:dynamicalDipolTransparencyNSI}
	\end{eqnarray}
	Using $\hat P_1$ to evaluate the dipole elements based on Eq.~\eqref{eq:dynamicalDipolElementsSymmetrySI}, we find
	$V_{\mu ,\mu' }^{(n)} = \alpha_V^{(P_2)}   e^{-in\Omega  t_{P_2}   }  V_{\mu',\mu }^{(n)*} $. Combining both results, we obtain
	\begin{equation}
	V_{\mu',\mu }^{(n)} = V_{\mu,\mu' }^{(n) } \alpha_V^{(P_2)}  \alpha_V^{(P_1)} e^{-in\Omega (t_{P_1} - t_{P_2})   }.
	\label{eq:symmetryDipoleElmentTransparencySI}
	\end{equation}
	Inserting Eq.~\eqref{eq:symmetryDipoleElmentTransparencySI} into  Eq.~\eqref{eq:transparencyRequirementNSI}, we find the transparency condition
	\begin{equation}
	\tilde	\chi_n(m\Omega) \propto 
	\begin{cases}
	0   &\text{if}\; e^{-in\Omega (t_{P_1} - t_{P_2})   }  = 1; \epsilon_\mu =\epsilon_{\mu'}=0 \\ 
	1  & \text{else} ,
	\end{cases}
	\label{eq:symmetryProtectedTransparencyConditionSI}
	\end{equation}
	which is the condition presented in the main text. Noteworthy, for  $e^{-in\Omega (t_{P_1} - t_{P_2})   } = -1$ the two resonances can also add up constructively.

	For illustration,  we consider the ac-driven two-level system (TLS) which is introduced in the main text in Eq.~(12). Its Hamiltonian reads
	\begin{align}
		\hat H_0(t)&= \frac{h_{\rm x} }{2}\hat \sigma_{\rm x} + \frac{f_{\Omega }}{2}  \cos\left(\Omega t \right) \hat \sigma_{\rm z} , 
		\label{eq:HamiltonianSI}
	\end{align}
	where $\hat \sigma_x,\hat \sigma_{\rm z}$ are the Pauli matrices, $h_x$ is the tunneling amplitude, and $f_{\Omega}$ the driving strength. The dipole transition operator  is $\hat V = \hat  \sigma_x$.

	For $h_{\rm x} =0$, a particle-hole symmetry is defined for $\hat P_1 = \hat \sigma_{\rm z} $ and $t_{P_1} = \tau/2$, and  a second particle-hole symmetry is given for $\hat P_2 = \mathbbm 1  $ and $t_{P_2} = \tau/2$. 
	The quasienergy spectrum is  flat, i.e., $\epsilon_\mu =0$, as depicted in Fig.~\ref{figSpectraSpinBosonSI}(a).
	As  $ \hat P_i \hat \sigma_x^{*}  \hat P_i = (-1)^i \hat  \sigma_x  $,  a symmetry-protected transparency with $\tilde	\chi_n(m\Omega) = 0$ appears  for all $n$ according to the condition Eq.~\eqref{eq:symmetryProtectedTransparencyConditionSI}. We note that the transitions for even $m$, e.g. $\hat	V_{\nu , \mu} ^{(-2)}\hat	V_{\mu,\nu } ^{(2)}$, are additionally suppressed because of a dark state selection rule imposed by the dynamical parity symmetry introduced in the letter.

	For a finite but small $h_{\rm x} \neq 0$, the symmetry $\hat P_2 $ is slightly broken. 
	The quasienergy spectrum is not flat anymore as can be seen in  Fig.~\ref{figSpectraSpinBosonSI}(b). There is still a quasienergy crossing close to $f_{\Omega} \approx 2.4\Omega$, giving rise to a coherent destruction of tunneling.
	In Fig.~\ref{figSpectraSpinBosonSI}(d) we depict $\tilde \chi_0(\omega_{\rm p})$. Because of the lifted quasienergy degeneracy, the transitions $\hat	V_{\nu , \mu} ^{(-1)}\hat	V_{\mu,\nu } ^{(1)}$ are now visible. However, the system is still transparent at the quasienergy crossing. This is a consequence of the symmetry-protected transparency, which  remains  approximately valid as the symmetry $\hat P_2 $ is only slightly broken. In order to emphasize the relation to the symmetry-protected transparency, we coin this weaker effect symmetry-induced transparency. 
	The symmetry-induced transparency scales as $\tilde	\chi_n(m\Omega)\propto h_x / \Omega $ at the crossing. It is related to a small asymmetry of the response peaks of $\tilde \chi_n(\omega_p)$. Instead of the Floquet-Gibbs distribution, we choose  $p_0= 0.6$ and $p_1 = 0.4$ as the stationary state distribution in Fig.~\ref{figSpectraSpinBosonSI}(c) to highlight the symmetry-induced transparency.
	
	Please note that there is additionally a dynamical parity symmetry of the TLS in Eq.~\eqref{eq:HamiltonianSI} for $\hat R = \sigma_{\rm x}$ and $t_R = \tau/2$. Consequently, the quantum number of the Floquet states related to the parity symmetry operation are $m_0=0$ and $m_1=1$.  According to the spDS condition Eq.~\eqref{eq:symmetryProtectedDarkStatesSI} for $N=2$ and $\alpha_V^{(R)} =1$, we find that $V_{0,1 }^{(n)}=V_{1,0 }^{(n)}= 0$ for even $n$.
	
\end{widetext}

\end{document}